          [2001/04/25 1.1 (PWD)]
\documentclass[a4paper,twocolumn]{esapub} 
\usepackage{natbib}
\usepackage{graphicx}

\def\gray {$\gamma$-ray\ }

\def\deg{^{\circ}}
\def\spidiffit{{\it spidiffit\ }}

\title{SPI measurements of  the diffuse Galactic hard X-ray continuum  }

\author{Andrew W. Strong }
\author{R. Diehl}
\author{H. Halloin}
\author{V. Sch\"onfelder}
 \affil{ Max-Planck-Institut f\"ur extraterrestrische Physik,   Postfach 1312, D-85741 Garching, Germany }
\author{L. Bouchet}
\author{P. Mandrou}
\affil{ Centre d'Etude Spatiale des Rayonnements, CNRS/UPS, B.P.4346, 31028 Toulouse Cedex 4, France}
\author{\\ B. J. Teegarden}
\affil{Code 661, NASA/Goddard Space Flight Center, Greenbelt, MD 20771, USA}
\author{F. Lebrun}
\affil{CEA-Saclay, DAPNI/Service d'Astrophysique, F91191 Gif sur Yvette Cedex, France}
\author{R. Terrier}
\affil{Astroparticules et Cosmologie, CNRS/Paris
7, 75231 Paris Cedex 5}

\begin{document}


\maketitle

\begin{abstract}
 INTEGRAL Spectrometer SPI data from the first year of the Galactic
 Centre Deep Exposure has been analysed for  the diffuse  continuum
 from the Galactic ridge.  A new catalogue of sources from the INTEGRAL
Imager IBIS has been used to account for their contribution to the
celestial signal.  Apparently diffuse emission is detected at a level
$\sim10\%$ of the total source flux. A comparison of the spectrum of diffuse emission with that from
an analysis of IBIS data alone shows that they are consistent.  The question of the contribution of
unresolved sources to this ridge emission is still open.
\end{abstract}
\section{Introduction}
The inner Galactic ridge  is known to be an intense source of 
continuum hard X- and soft \gray emission. The hard X-ray emission was
discovered in 1972 \citep{bleach72} and has subsequently been observed
from keV to MeV energies by ASCA, Ginga, RXTE, OSSE, COMPTEL and most
recently by Chandra and XMM-Newton.

  While the physical process ($e^+e^-$ annihilation) producing the
  positron line and positronium continuum is clear, for of the
  remaining continuum it is not, although nonthermal bremsstrahlung is
most likely \citep{dogiel02a};
 the implied photon luminosity in the continuum is a few 10$^{38}$ erg s$^{-1}$.
 Its origin in a point-source population seems
unlikely at keV energies \citep{tanaka99} since there are no known candidate objects with the required spatial and luminosity properties.
In addition, high-resolution imaging in X-rays with Chandra \citep{ebisawa01} shows a truly diffuse component.
 Recently \citet{hands04} have used the XMM-Newton Galactic Plane
 Survey to show that, in the 2--10 keV band, 80\% of the ridge emission (l=19$\deg$-22$\deg$)
 is probably diffuse, and only 9\% can be
accounted for by Galactic sources (the rest being extragalactic
sources).

   Non-thermal \gray emission in the interstellar medium implies a
very high luminosity in energetic particles \citep{dogiel02a}.
Emission mechanisms have been discussed by \citet{valinia00a,valinia00b} and
 \citet{tanaka02}, and a solution to the  problem of energetics has been
 suggested by \citet{dogiel02b} but the issue is quite open.
\citet{valinia00b} propose a composite model with thermal and nonthermal components from electrons accelerated in supernovae or ambient interstellar turbulence.
 At MeV energies the origin of the emission is also uncertain \citep{SMR00}. The study
 of this emission is a key goal of the INTEGRAL mission.  The high
 spectral resolution combined with its imaging capabilities lead to
high expectations for investigating the nature of this enigmatic
emission.

Apart from its intrinsic interest,  reliable modelling of the diffuse emission
will be essential for the study of point sources in the inner Galaxy,
since it  contributes a large anisotropic background against which they must be observed.

The most directly comparable results to ours (with respect to energy range) 
 are from OSSE on the COMPTON
Gamma Ray Observatory. \citet{purcell96} used coordinated observations of sources in the Galactic centre region from the SIGMA satellite to evaluate the source contribution. They concluded that after correction for these sources,
the OSSE flux was consistent with that measured by OSSE at
 l=25$\deg$ and 339$\deg$.
\citet{kinzer99,kinzer01} give an OSSE spectrum for $l=b=0$,  but without independent imaging observations, so that the point-source contribution remains uncertain.

A preliminary analysis of diffuse continuum emission using SPI data has
been given in \citet{strong03}, using the first cycle of INTEGRAL GCDE
observations.  It was concluded that the diffuse emission was detected
at a level consistent with previous experiments, but the systematic
errors were large, due in part to the uncertainty in the point-source
contribution.  Only four of  the strongest point sources were included
in that analysis, since at least at higher energies ($>$100 keV) the
source contribution is not thought to be critical. At lower energies an
adequate \gray source catalogue (e.g. from INTEGRAL) was not yet available, and it was stated in \citet{strong03} that
 point-source contamination was likely. Note that the adopted method of
analysis  has the property that sources which are unaccounted
for are mainly `absorbed' into the fitted instrumental background
rather than affecting the diffuse signal.

Meanwhile, a new analysis of IBIS data \citep{lebrun04,terrier04} has
shown that indeed a large fraction of the total \gray emission from the inner
Galaxy is due to sources, at least up to 100 keV.  This work has
produced a source catalogue containing 91 sources, which can now be
used as input to the SPI model fitting, giving a much more solid basis for
the analysis.  This exploits the complementarity of the  instruments on
INTEGRAL for the first time in the context of diffuse emission.

\section{Data}
The INTEGRAL Core Program \citep{winkler03} includes the Galactic Centre Deep Exposure (GCDE)
which maps the inner Galaxy ($330\deg<l<30\deg, -20\deg<b<20\deg$) with
a viewing time of about 4Ms per year. The full region is covered in one GCDE cycle,
and there are two cycles per year. Data from the first two cycles are used for the study   reported here. 

We use data from the SPI (INTEGRAL Spectrometer) instrument;
descriptions of the instrument and performance are given in \citet{vedrenne03,attie03}.  The energy range covered by
SPI is 20~keV -- 8~MeV, but here we restrict the analysis to energies
up to 1 MeV; above these energies the statistics are small and the
analysis is more difficult, so is reserved for future work.  
The data were pre-processed using the INTEGRAL Science Data Centre (ISDC) Standard Analysis   software (OSA) up to
the level containing binned events, pointing and livetime information.
GCDE data from 33 orbital revolutions from  47 -- 123 were used.  2348 pointings were used,  as shown in Fig 1; the total  exposure livetime is $3.34\times 10^6$ s. 
The exposure per pointing is typically 1800s, but can
be as low as 300s in cases of high telemetry losses in the early part of the
 GCDE.  The energy calibration is performed using instrumental
background lines with known energies; while this is a critical operation
for line studies (where sub-keV accuracy is required), for continuum
studies a standard calibration ($\sim$1 keV accuracy) is quite adequate.
Various energy binnings were used, depending on the available statistics as a function of energy.
  Only single-detector events are used here.

 The instrumental response is based on extensive Monte Carlo simulations and
parameterization \citep{sturner03}; this has been
tested on the Crab in-flight calibration observations and shown to be
reliable to better than  20\% in absolute flux at the current state of the analysis.

Since INTEGRAL data are dominated by instrumental background, the
analysis needs to have good background treatment methods. In the present
work the background ratios between detectors are  obtained by averaging the entire
dataset over time; this has been found to give results which hardly differ from
taking `OFF' observations, and has the advantage of much higher statistics (few OFF data are available at present). A check on the systematics in this  background approach was given in \citet{strong03}.

\section{Method}
 We use the program \spidiffit which fits the data to a linear
combination of input astronomical skymaps (e.g. HI and CO surveys,
emission models) and point sources, together with background
components. The fit is performed by maximum-likelihood with one
parameter per component and energy range. The background is fitted per
pointing using a template for the ratios between detectors from the average over all observations. Hence the time-dependence of the background is explicitly
determined from the data themselves on the assumption of constant
detector ratios.

Since the distribution of the emission is unknown but certainly
correlated with tracers of large-scale Galactic structure and the diffuse interstellar medium, we include
line-of-sight integrated HI and CO surveys (which trace the atomic and
molecular gas) as basis models.
 In addition, since the positronium emission is an important source of
continuum below 511 keV and is believed to include in addition a
component more concentrated towards the Galactic centre than the gas
tracers \citep{kinzer01,milne02,jean03,knodlseder03}, and noting that
\citet{kinzer01} and \citet{milne02} found no significant difference
between the line and continuum annihiliation angular distributions,
 we include also a Gaussian with FWHM 10$\deg$ centred on $l = b = 0$.
   Note that each fitted component has an independent spectrum so that
   it can be separated in the fitting; the sum of the diffuse
components then gives the total diffuse emission spectrum of the  inner
Galaxy.  Additionally we remark that a coded-mask telescope is only
sensitive to flux contrasts, so isotropic emission is suppressed.

For the sources we use the catalogue  from the IBIS analysis described in
\citet{lebrun04,terrier04} and Terrier (private communication). This contains 91 sources;
 only the source positions are required since the source spectra are determined in the model fitting.
  
\section{Results}
Fig 2 illustrates the temporal variation of the background scaling factor as
determined in the fitting, for the energy range 18-178 keV. Other
energy ranges appear very similar. The increase between the 1st and 2nd
GCDEs is clearly visible around pointing sequence 1600. High points
represent increases due to end-of-orbit passes  or other physical
effects. Generally the fluctuations are at the few percent level and
show that the technique effectively determines the time-dependence of the
background. Alternative background treatments using predictions based on
activation tracers \citep{jean03} provide consistent background histories.

 Fig 3 shows the spectrum of total emission from the 3
diffuse components, and the total from the sources included in the
fitting. 
 The diffuse
emission is approximately 10\% of the summed sources, consistent with
\citet{lebrun04,terrier04}. It is slightly lower than given in \citet{strong03}
due to the additional sources included, but still within the error bars
quoted in that paper.  The positronium edge is detected,  at low
significance.  (Note that the broad bins used here to study the
continuum suppress the 511 keV line).
 No diffuse emission is detected above 500 keV, but this is still
consistent within the error bars with the COMPTEL fluxes.

 For comparison we show in Fig 4 the scaled (l=b=0) OSSE
spectrum  and results from RXTE
\citep{revnivtsev03} and COMPTEL \citep{strong99}. 
The OSSE fluxes are scaled as described in \citet{strong03} (continuum scaled by 0.5, positronium scaled by 0.3), since a precise comparison is difficult because of the different instrument responses.  A scaling factor less than unity is expected
between the l=b=0 OSSE  flux per radian and the full inner radian flux measured by SPI because of the concentration of the emission towards the Galactic centre.
The RXTE results for $|b|>2^o$ are scaled to the SPI region using information provided by M. Revnivtsev (private communication).
\section{Discussion}
Detailed quantitative comparison with previous measurement of the diffuse continuum emission (from OSSE, RXTE, COMPTEL) is not
possible because of the different sky areas and energy ranges covered,
and the very different instrumental characteristics. Hence the
comparison shown in Fig 4 is illustrative only, but shows that the
INTEGRAL results are compatible with or somewhat lower than found in previous work. 
The fact that we
can now explicitly include a large number of sources in the analysis makes it 
more robust.

The question remains as to whether the apparently diffuse
emission is really interstellar or is the superposition of sources
below the present detection threshold. This issue can only be answered
in the future with population synthesis models and studying the
logN-logS of the detected sources.  We note however that at energies
just below those visible to INTEGRAL, similar issues are now being
addressed by Chandra and XMM-Newton (see  Introduction).  The
finding of \citet{hands04} with XMM that $\sim$80\% of the Galactic
2--10 keV emission is diffuse supports the RXTE analysis of
\cite{revnivtsev03}. Hence we do expect  diffuse
emission at some level in the INTEGRAL range unless a dramatic cutoff occurs over a
factor of a few in energy.  

The apparent difference in the diffuse-to-source ratio between X-ray and gamma-ray energies can probably be understood as a sampling effect: while the present gamma-ray ratio includes all the bright inner Galaxy  sources, the XMM ratio mentioned above is for a small region which does not sample the brightest sources.
Using  RXTE scan observations, the result of \citet{ebisawa99}  shows that for a wide latitude region the X-ray keV diffuse-to-source ratio is  comparable with what we find for the inner Galaxy.

Evaluation of the combined results from
X-ray and gamma-ray instruments will be essential and will indeed be
possible in the near future.

In Fig 5 we compare the summed source spectra with the MeV ridge emission
measured by COMPTEL  \citep{strong99}. This shows that a significant
contribution to the 1 - 30 MeV emission could be due to the extension
of the source spectra into this range, consistent with the  requirement for an unresolved source component suggested by \citet{SMR00}.

\section{Comparison with IBIS results}
Fig 6 compares the diffuse spectrum from this work with that derived
using INTEGRAL/IBIS ISGRI data by \citet{terrier04} (see Introduction). 
The results are  consistent,
both in the low-energy range (20 -- 40 keV) , where the IBIS analysis gives  an excess not accounted for
by detected sources, and at higher energies (40 -- 220 keV) where IBIS was only able to derive upper limits.

\section{Conclusions}
We have exploited the complementarity of the instruments on INTEGRAL
to improve the SPI analysis of diffuse Galactic \gray emission from 20 -1000 keV.
Inclusion of the sources found by IBIS improves the sensitivity and reliability of the analysis. Diffuse emission is detected at a level rather lower than previously when the IBIS sources are accounted for. However this marks only the beginning of what will be possible by jointly analysing the data from both INTEGRAL instruments with their unique combination of characteristics. 
\begin{figure}
\centering
\includegraphics[width=1.05\linewidth]{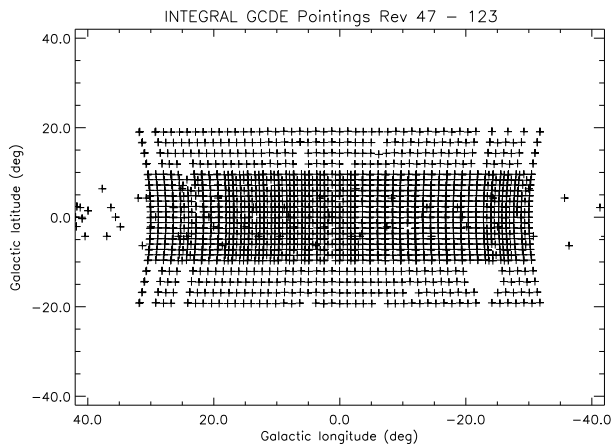}
\caption{INTEGRAL pointings used for this analysis. 
\label{fig:single}}
\end{figure}
\begin{figure}
\centering
\includegraphics[width=1.05\linewidth]{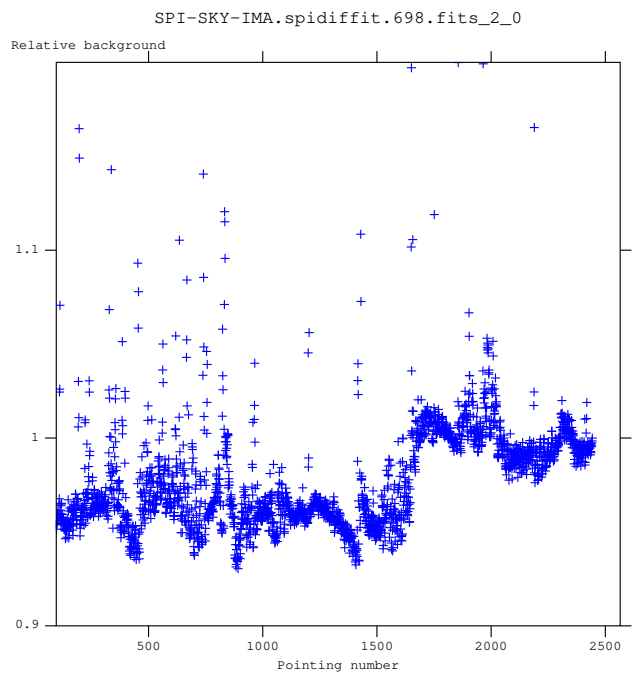}
\caption{Instrumental background scaling factor determined by the fitting procedure. 
Note that the pointing number is sequential, and there is a time gap between the 1st and 2nd GCDEs around pointing 1600.
\label{fig:single}}
\end{figure}

\begin{figure}
\centering
\includegraphics[width=1.05\linewidth]{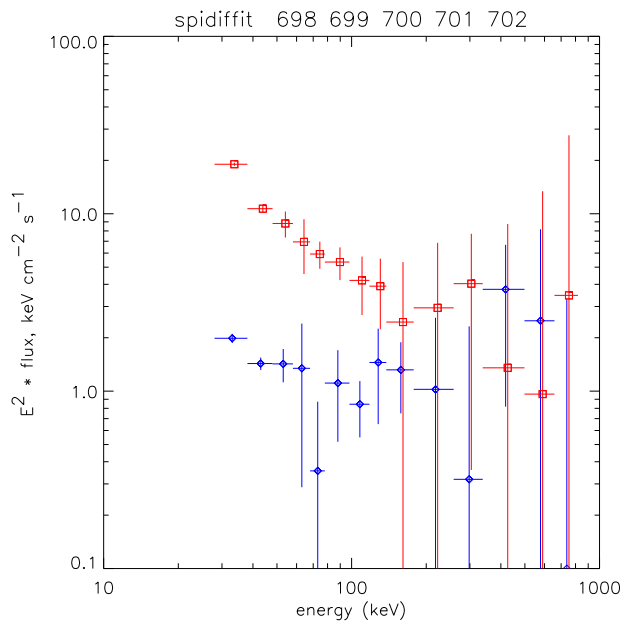}
\caption{Diffuse continuum (blue diamonds) and summed source (red squares) spectra, from this SPI analysis. The diffuse emission is the sum of the three fitted model  components.
 \label{fig:single}}
\end{figure}
\begin{figure}
\centering
\includegraphics[width=1.05\linewidth]{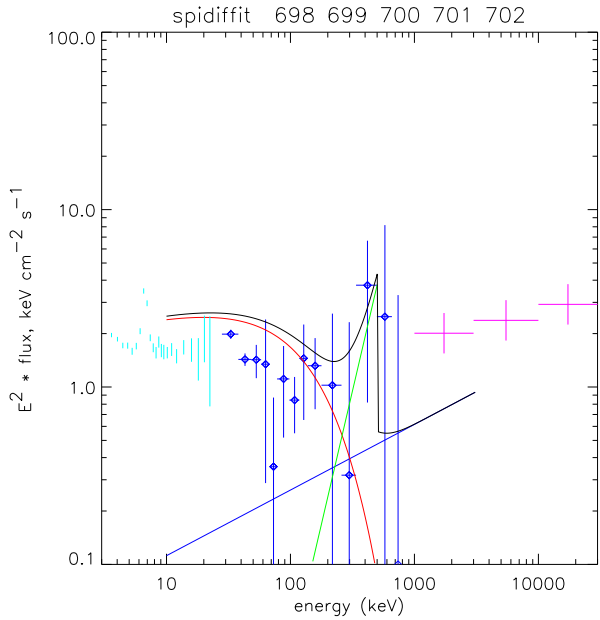}
\caption{Diffuse continuum from this SPI  analysis (blue diamonds). 
  Comparison with  with 1.
 OSSE spectrum around $l = b = 0$ from \citet{kinzer99},
 with components:  exponentially cutoff power law (red), high-energy continuum power law (blue), positronium (green) and total (black). Components are  scaled as  in \cite{strong03} and described in the text. 
2. RXTE from \cite{revnivtsev03} and Revnivtsev (private communication) (light blue bars), 3. COMPTEL from \cite{strong99} (magenta). 
\label{fig:single}}
\end{figure}
\begin{figure}
\centering
\includegraphics[width=1.05\linewidth]{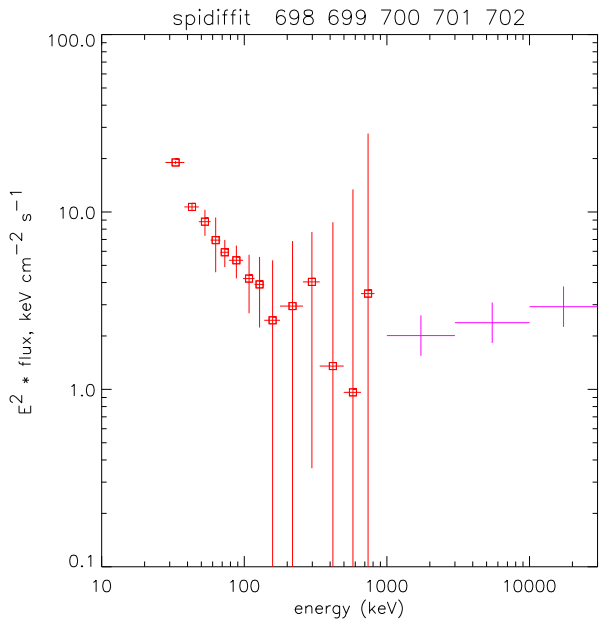}
\caption{
  Summed SPI source spectra from this analysis (red squares) compared with ridge emission measured by   COMPTEL from \cite{strong99} (magenta). 
\label{fig:single}}
\end{figure}
\begin{figure}
\centering
\includegraphics[width=1.05\linewidth]{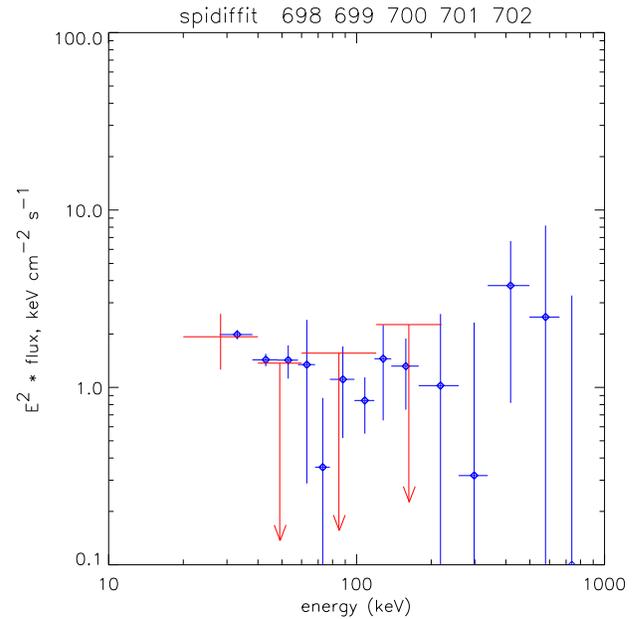}
\caption{
  Diffuse Ridge emission measured by  SPI (this work): dark blue, compared with results from  IBIS \citep{terrier04}: red. 
\label{fig:single}}
\end{figure}

\section*{Acknowledgments}
We thank Mikhail Revnivtsev and Ken Ebisawa for useful discussions.

\end{document}